# Nearby Candidate Dust-Disk Pre-Main-Sequence Solar-Mass Stars

*Anatoly A. Suchkov* (*STScI, 3700 San Martin Dr. Baltimore, MD 21218*)

**Abstract.** I have isolated a population of numerous F stars that appear to be pre-main-sequence (PMS). The candidate PMS stars have been identified using CM diagram, reddening, flux excess in the UV and near-infrared, and luminosity anomaly. Negative luminosity anomaly and excessive UV flux for many of these stars is suggestive of accretion disks, while the NIR excess is indicative in many cases of the presence of dust disk thermal emission. Observed overluminosity of many PMS candidates is consistent with their pre-main-sequence status. The bulk of the PMS candidates is located within 200 pc, exhibiting strong association with regions of star formation that are numerous between ~130 to 180 pc. The number of PMS candidates plummets redward of the spectral type ~ F5. This effect may provide important clues for understanding the evolution of PMS stars in the solar-mass range.

**Introduction.** Formation of a planetary system around a star is a much more conspicuous phenomenon than any manifestation of planets themselves, thus it can be detected and studied more easily than extrasolar planets. Since it is believed to occur simultaneously with or right after the star formation event, very young and pre-main-sequence (PMS) stars are the most popular targets in searches for signatures of extrasolar planetary systems. Obviously, of all young stars, the solar-type stars are of a special interest, because they can provide a clue as to whether planetary systems are a regular attribute of stars similar to the Sun. In this paper we report the detection of a rather large population of stars, only slightly more massive than the Sun, that appear to be mostly PMS, many probably with dust disks signifying recent or ongoing planetary system formation. The stars discussed here are from the sample of ~12000 F stars that have both the Hipparcos and *uvby* data (Suchkov & McMaster 1999, Suchkov 2001), about half of them with the 2MASS *J, H,* and *K* magnitudes (Suchkov et al. 2002).

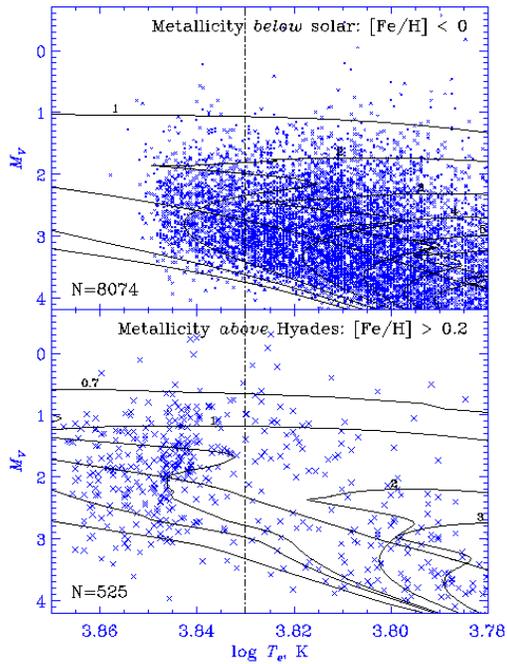
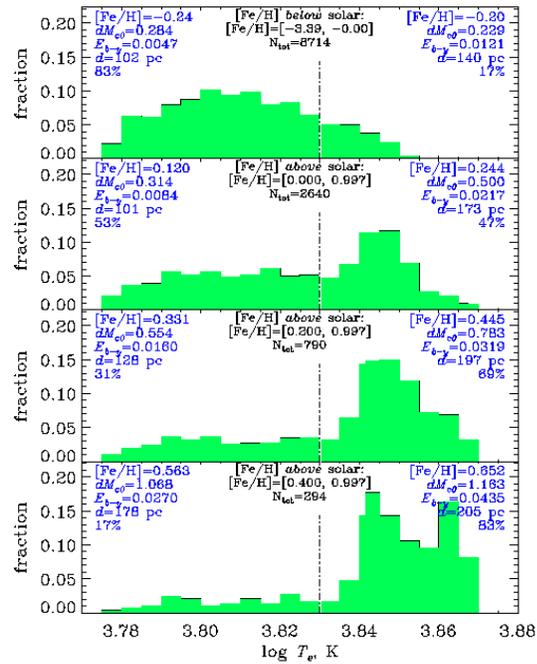

Figure 1. log $T_e$ - $M_V$ diagram for two metallicity groups of F stars.

Figure 2. Temperature distribution of F stars as dependent on metallicity.

**PMS stars from the log Te – MV diagram.** PMS stars are expected to be metal rich, therefore the first place to look for them is among stars with high metallicities. Figure 1 reveals a remarkable difference between the two metallicity groups of F stars. At metallicities below solar, the fraction of stars decreases toward high temperatures in the temperature range log $T_e$ > 3.80, which is expected because of the declining IMF and shorter MS lifetimes toward massive, hence hotter stars (temperature distribution at $T_e$ < 3.80 for our sample is dominated by sample selection effects). But the metal rich stars exhibit strong concentration at high rather than low temperatures, log $T_e$ > 3.83 (spectral types earlier than ~F5). That behavior is incompatible with the two indicated factors that drive the fraction of stars down at high temperatures. This suggests that most of the metal rich stars at the 'hot' end of the log $T_e$–$M_V$ diagram are not regular MS or post-MS stars. The alternative is that they are dominated by PMS stars. Their large luminosity spread, quite conspicuous in Figure 1, is in perfect agreement with the PMS hypothesis. Indeed, if the hot stars above the ZAMS in the lower panel of Figure 1 were normally evolving post-MS stars, the diagram would have been dominated by stars in the low- rather than high-temperature range.

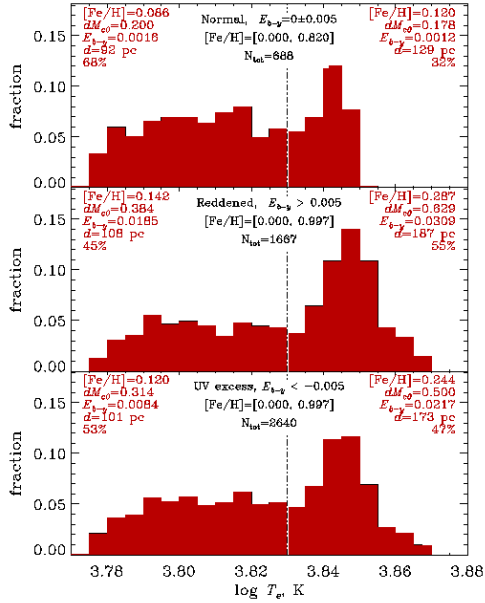
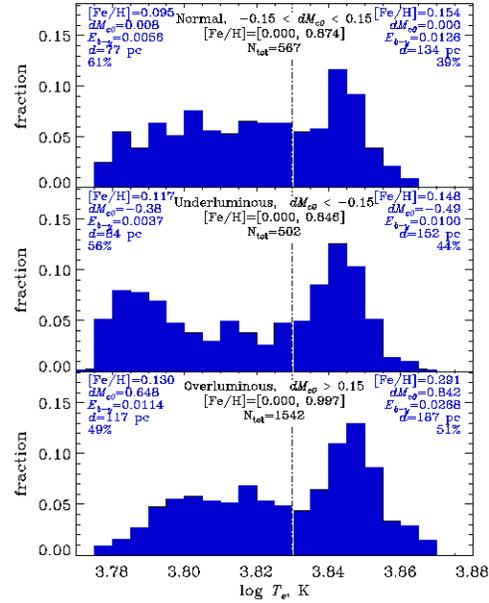

Figure 3. Temperature distribution of F stars as dependent on the reddening parameter.

Figure 4. Temperature distribution of F stars as dependent on luminosity anomaly.

The discrepancy between the expected and the actual temperature distribution for metal rich stars gets larger with increasing metallicity (see Figure 2; the legend on the right and the left in this and the next two Figures, 3 and 4, give the mean value of respective parameters for stars cooler and hotter than log $T_e$ = 3.83, respectively). That is exactly what one would anticipate if the stars are mostly PMS, because the fraction of such stars should obviously be the largest among most metal rich stars.

Within the context of PMS stars, it is especially interesting that the observed distribution of the PMS candidates plummets redward of the spectral type ~F5. That effect may provide important clues for understanding the evolution of solar-mass PMS stars.

**PMS stars from the reddening parameter.** The temperature distribution of metal rich stars in Figure 3 shows that both reddened stars (middle panel) and those with anomalous (negative) reddening parameter (lower panel) are more numerous at high temperatures than unreddend stars (upper panel). That suggests that association of metal rich stars with PMS stars is more likely for the first two groups. For the PMS reddened stars, $E_{b-y}$ > 0, this is to be expected because of circumstellar dust and a general dusty star formation environment. On the other hand, intrinsically negative values of the reddening parameter as derived from the *uvby* photometry, $E_{b-y}$ < 0, imply a blue excess and thus suggest the presence of an associated UV excess. Some PMS stars are known to be excessive UV emitters. The extra UV emission is believed to be generated as a result of accretion of a circumstellar disk material onto a star (e.g., Gullbring et al. 1998, Rebull et al. 2002). Therefore, hot stars with anomalous reddening (lower panel) are strong candidates for being PMS with accretion disks.

**PMS stars from anomalous luminosity.** For a normally evolving F star, the amount of UV emission blueward of the Balmer jump (band *u* in the *uvby* system) is sensitive to stellar surface gravity. This is used to predict stellar luminosity based on *uvby* colors. Any anomaly in the UV flux of a star would result in a discrepancy between its predicted and actual absolute magnitudes, $M_{c0}$ and $M_V$, respectively. A star with excessive UV emission would look underluminous for its UV flux, $dM_{c0} = M_{c0} - M_V < 0$. As mentioned above, extra UV luminosity in young stars is probably due to the accretion of the circumstellar disk material. If so, the underluminous hot stars displayed in the middle panel of Figure 4 are obvious PMS candidates with accretion disks.

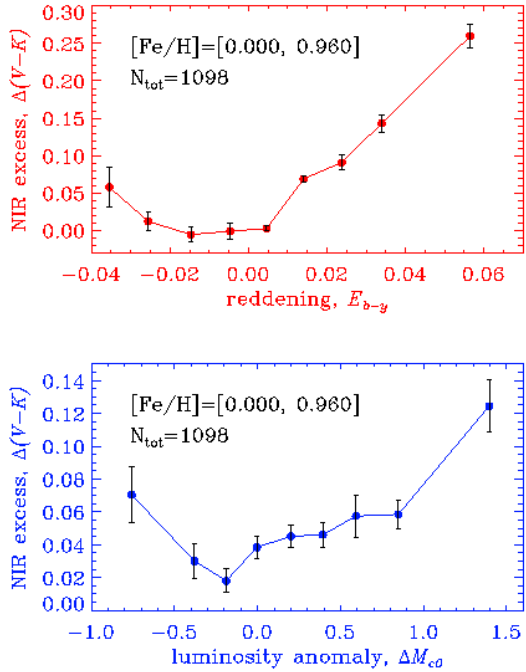

Figure 5. The NIR excess of F stars as a function of reddening (upper panel) and luminosity anomaly (lower panel).

Overluminosity of the stars in the lower panel of Figure 4 is partly associated with the presence of unidentified binaries with comparably bright components (Suchkov & McMaster 1999). In general, however, many overluminous stars are intrinsically too luminous for their UV flux. In the case of old stars, this seems to be a result of anomalous post-main-sequence stellar evolution (Suchkov 2001). However, for the reasons discussed above, post-MS stars cannot make any significant contribution to the total number of the hot stars in the lower panel of Figure 4. But PMS stars can. Their overluminosity should be expected in the case of weak or no UV excess (hence no strong accretion) because of structural differences between pre- and post-main-sequence stars (no helium core in PMS stars, hence surface gravity different, at the same temperature and luminosity, than that in post-MS stars). Thus, underluminous and overluminous metal rich stars displayed in Figure 4 provide a sample of PMS candidates with and without effects of strong accretion, respectively.

**Accretion disks from the NIR excess, reddening, and luminosity anomaly.** Figure 5 shows the behavior of the near-infrared excess, $\Delta(V-K)_{b-y}$ (Suchkov et al. 2002), of metal rich F stars as a function of the reddening parameter, $E_{b-y}$, and luminosity anomaly, $dM_{c0}$. In the range of positive values of $E_{b-y}$, the increase of the NIR excess toward higher reddening is due to interstellar and/or circumstellar reddening *per se* plus possible thermal emission of a circumstellar dust disk (Suchkov et al. 2002). In the range of negative values, $E_{b-y} < 0$, the NIR excess trends higher as the reddening parameter decreases. As mentioned above, the reddening parameter gets negative if a star emits excessively in the UV, a particular cause of such an excess being disk accretion. Therefore, the trend is suggestive of accretion disks (enhanced stellar UV) that emit thermal NIR stronger at higher accretion rates. This inference is consistent with the presence of a similar trend in the lower panel of Figure 5 (at $dM_{c0} < 0$, the NIR excess gets higher with decreasing $dM_{c0}$). Indeed, since excessive UV emission drives $dM_{c0}$ down, larger near-infrared excess at large negative values of luminosity anomaly is exactly what one would expect for accretion-disk stars in which the dust NIR emission correlates with the enhanced stellar UV emission.

In the range $dM_{c0} > 0$, the larger luminosity anomaly, the farther away from the ZAMS the younger the putative PMS stars are. Since younger systems are, at a given temperature, more luminous and typically possess more circumstellar dust, they are expected to have a stronger NIR emission. Therefore, as long as UV excess is small enough or absent, the thermal infrared emission should be stronger at higher $dM_{c0}$. The behavior of $\Delta(V-K)_{b-y}$ in the lower panel in Figure 5 at $dM_{c0} > 0$ is thus consistent with very dusty PMS stars that strongly absorb UV flux or have low disk accretion rates.

**Association with star formation regions.** Sky distribution of stars with and without near-infrared excess (Figure 6) reveals a remarkable difference between these two groups. The first group (left panel) exhibits strong association with at least two most famous regions of ongoing star formation, ρ Oph and Tau – Aur, that are centered at a distance of ~150 pc. At the same time there is nothing special about these regions in the distribution of stars in the second group (right panel). Thus, the PMS candidates identified in this paper on the basis of their NIR excess map nicely the known regions of star formation. The mapping is quite good not only in the sky but also along the line of sight: The distance distribution of the PMS candidates within the distance range 100 to 200 pc is peaked at ~150 pc, meaning that most of those stars are indeed spatially associated with the star-forming regions.

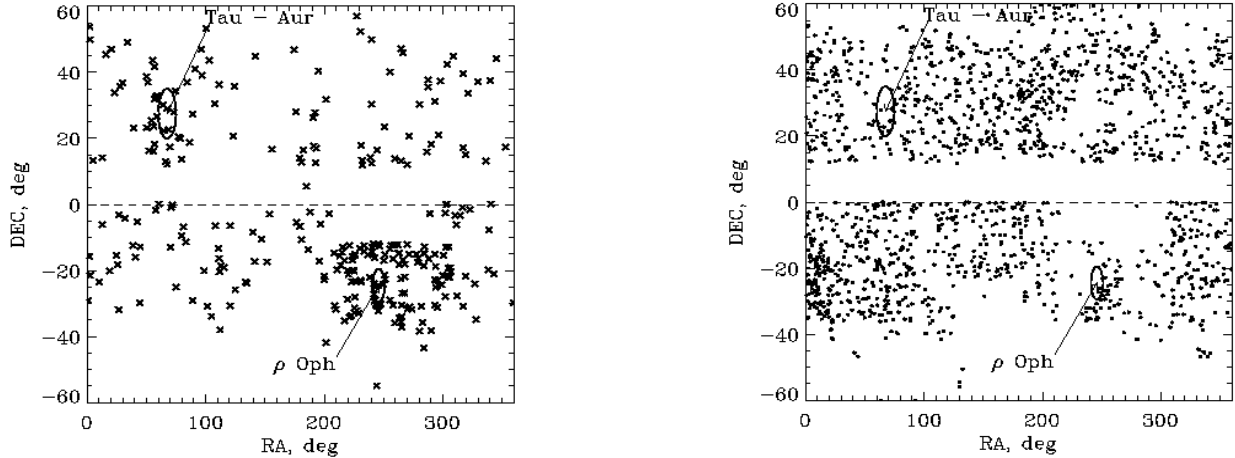

Figure 6. Sky distribution of stars within 100 to 200 pc with and without infrared excess, $\Delta(V-K)_{b-y} > 0.1$ (left) and $\Delta(V-K)_{b-y} < 0.1$ (right), respectively.

A more detailed analysis unveils many interesting features in that association. In particular, it appears that candidate PMS F stars envelope rather than get immersed into molecular clouds in the center of a star-forming region. This may be, of course, a result of an observational bias, because stars within or behind molecular clouds are hard to detect due to strong extinction. On the other hand, concentration of PMS F stars at the edges of star-forming regions may well be a real phenomenon, reflecting changes in the ISM accompanying star formation events. Further studies of the association of the candidate PMS F stars with molecular clouds in star-forming regions would certainly help to better understanding the history of these regions.

# Nearby Candidate Dust-Disk Pre-Main-Sequence Solar-Mass Stars

*Anatoly A. Suchkov* (*STScI, 3700 San Martin Dr. Baltimore, MD 21218*)

**Abstract.** I have isolated a population of numerous F stars that appear to be pre-main-sequence (PMS). The candidate PMS stars have been identified using CM diagram, reddening, flux excess in the UV and near-infrared, and luminosity anomaly. Negative luminosity anomaly and excessive UV flux for many of these stars is suggestive of accretion disks, while the NIR excess is indicative in many cases of the presence of dust disk thermal emission. Observed overluminosity of many PMS candidates is consistent with their pre-main-sequence status. The bulk of the PMS candidates is located within 200 pc, exhibiting strong association with regions of star formation that are numerous between ~130 to 180 pc. The number of PMS candidates plummets redward of the spectral type ~ F5. This effect may provide important clues for understanding the evolution of PMS stars in the solar-mass range.

**Introduction.** Formation of a planetary system around a star is a much more conspicuous phenomenon than any manifestation of planets themselves, thus it can be detected and studied more easily than extrasolar planets. Since it is believed to occur simultaneously with or right after the star formation event, very young and pre-main-sequence (PMS) stars are the most popular targets in searches for signatures of extrasolar planetary systems. Obviously, of all young stars, the solar-type stars are of a special interest, because they can provide a clue as to whether planetary systems are a regular attribute of stars similar to the Sun. In this paper we report the detection of a rather large population of stars, only slightly more massive than the Sun, that appear to be mostly PMS, many probably with dust disks signifying recent or ongoing planetary system formation. The stars discussed here are from the sample of ~12000 F stars that have both the Hipparcos and *uvby* data (Suchkov & McMaster 1999, Suchkov 2001), about half of them with the 2MASS *J, H,* and *K* magnitudes (Suchkov et al. 2002).

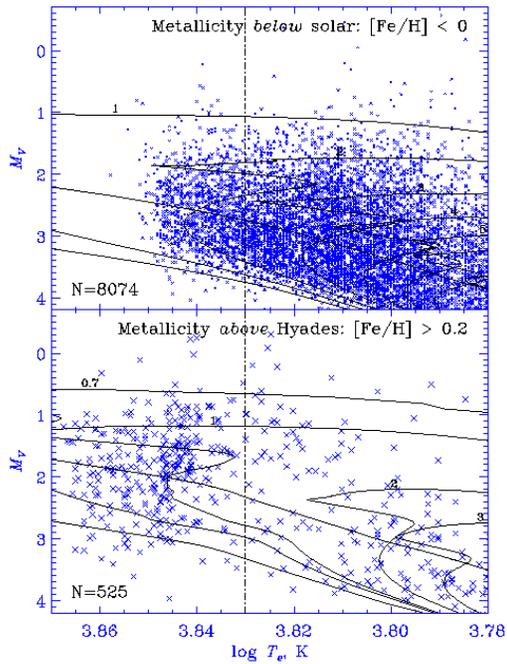
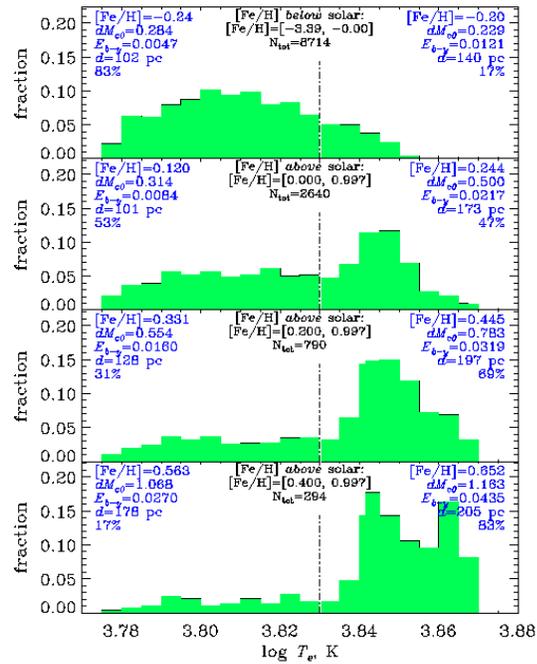

Figure 1. log $T_e$ - $M_V$ diagram for two metallicity groups of F stars.

Figure 2. Temperature distribution of F stars as dependent on metallicity.

**PMS stars from the log Te – MV diagram.** PMS stars are expected to be metal rich, therefore the first place to look for them is among stars with high metallicities. Figure 1 reveals a remarkable difference between the two metallicity groups of F stars. At metallicities below solar, the fraction of stars decreases toward high temperatures in the temperature range log $T_e$ > 3.80, which is expected because of the declining IMF and shorter MS lifetimes toward massive, hence hotter stars (temperature distribution at $T_e < 3.80$ for our sample is dominated by sample selection effects). But the metal rich stars exhibit strong concentration at high rather than low temperatures, $\log T_e > 3.83$ (spectral types earlier than ~F5). That behavior is incompatible with the two indicated factors that drive the fraction of stars down at high temperatures. This suggests that most of the metal rich stars at the 'hot' end of the $\log T_e$–$M_V$ diagram are not regular MS or post-MS stars. The alternative is that they are dominated by PMS stars. Their large luminosity spread, quite conspicuous in Figure 1, is in perfect agreement with the PMS hypothesis. Indeed, if the hot stars above the ZAMS in the lower panel of Figure 1 were normally evolving post-MS stars, the diagram would have been dominated by stars in the low- rather than high-temperature range.

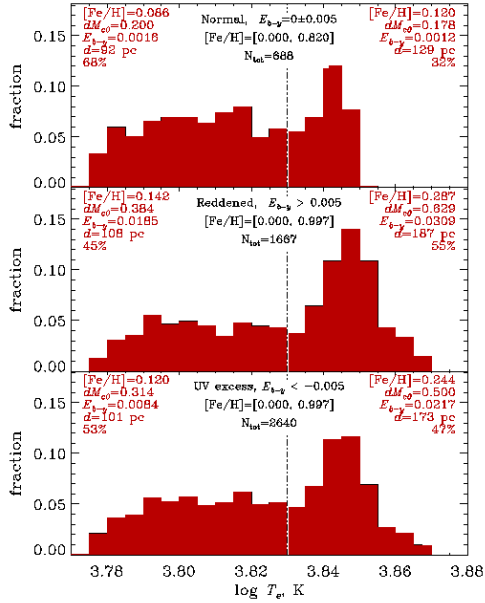
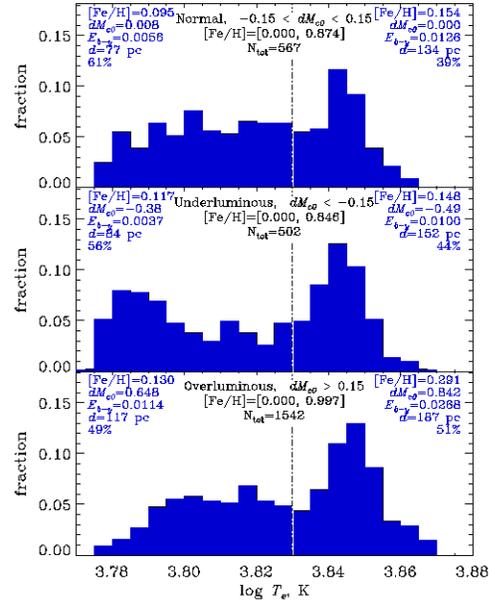

Figure 3. Temperature distribution of F stars as dependent on the reddening parameter.

Figure 4. Temperature distribution of F stars as dependent on luminosity anomaly.

The discrepancy between the expected and the actual temperature distribution for metal rich stars gets larger with increasing metallicity (see Figure 2; the legend on the right and the left in this and the next two Figures, 3 and 4, give the mean value of respective parameters for stars cooler and hotter than $\log T_e = 3.83$, respectively). That is exactly what one would anticipate if the stars are mostly PMS, because the fraction of such stars should obviously be the largest among most metal rich stars.

Within the context of PMS stars, it is especially interesting that the observed distribution of the PMS candidates plummets redward of the spectral type ~F5. That effect may provide important clues for understanding the evolution of solar-mass PMS stars.

**PMS stars from the reddening parameter.** The temperature distribution of metal rich stars in Figure 3 shows that both reddened stars (middle panel) and those with anomalous (negative) reddening parameter (lower panel) are more numerous at high temperatures than unreddend stars (upper panel). That suggests that association of metal rich stars with PMS stars is more likely for the first two groups. For the PMS reddened stars, $E_{b-y} > 0$, this is to be expected because of circumstellar dust and a general dusty star formation environment. On the other hand, intrinsically negative values of the reddening parameter as derived from the *uvby* photometry, $E_{b-y} < 0$, imply a blue excess and thus suggest the presence of an associated UV excess. Some PMS stars are known to be excessive UV emitters. The extra UV emission is believed to be generated as a result of accretion of a circumstellar disk material onto a star (e.g., Gullbring et al. 1998, Rebull et al. 2002). Therefore, hot stars with anomalous reddening (lower panel) are strong candidates for being PMS with accretion disks.

**PMS stars from anomalous luminosity.** For a normally evolving F star, the amount of UV emission blueward of the Balmer jump (band *u* in the *uvby* system) is sensitive to stellar surface gravity. This is used to predict stellar luminosity based on *uvby* colors. Any anomaly in the UV flux of a star would result in a discrepancy between its predicted and actual absolute magnitudes, $M_{c0}$ and $M_V$, respectively. A star with excessive UV emission would look underluminous for its UV flux, $dM_{c0} = M_{c0} - M_V < 0$. As mentioned above, extra UV luminosity in young stars is probably due to the accretion of the circumstellar disk material. If so, the underluminous hot stars displayed in the middle panel of Figure 4 are obvious PMS candidates with accretion disks.

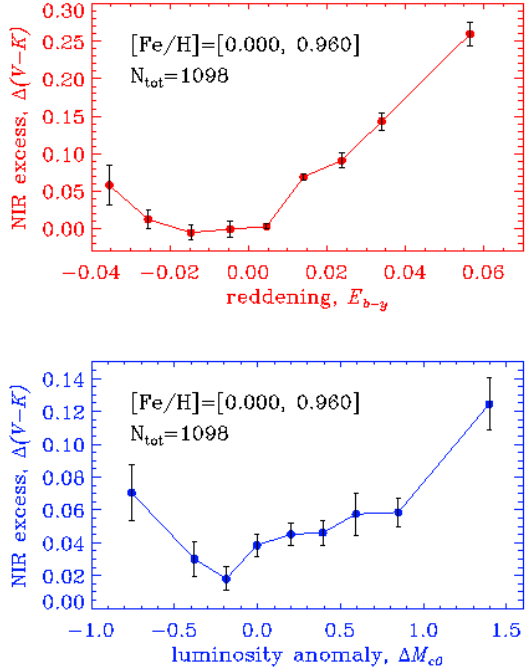

Figure 5. The NIR excess of F stars as a function of reddening (upper panel) and luminosity anomaly (lower panel).

Overluminosity of the stars in the lower panel of Figure 4 is partly associated with the presence of unidentified binaries with comparably bright components (Suchkov & McMaster 1999). In general, however, many overluminous stars are intrinsically too luminous for their UV flux. In the case of old stars, this seems to be a result of anomalous post-main-sequence stellar evolution (Suchkov 2001). However, for the reasons discussed above, post-MS stars cannot make any significant contribution to the total number of the hot stars in the lower panel of Figure 4. But PMS stars can. Their overluminosity should be expected in the case of weak or no UV excess (hence no strong accretion) because of structural differences between pre- and post-main-sequence stars (no helium core in PMS stars, hence surface gravity different, at the same temperature and luminosity, than that in post-MS stars). Thus, underluminous and overluminous metal rich stars displayed in Figure 4 provide a sample of PMS candidates with and without effects of strong accretion, respectively.

**Accretion disks from the NIR excess, reddening, and luminosity anomaly.** Figure 5 shows the behavior of the near-infrared excess, $\Delta(V-K)_{b-y}$ (Suchkov et al. 2002), of metal rich F stars as a function of the reddening parameter, $E_{b-y}$, and luminosity anomaly, $dM_{c0}$. In the range of positive values of $E_{b-y}$, the increase of the NIR excess toward higher reddening is due to interstellar and/or circumstellar reddening *per se* plus possible thermal emission of a circumstellar dust disk (Suchkov et al. 2002). In the range of negative values, $E_{b-y} < 0$, the NIR excess trends higher as the reddening parameter decreases. As mentioned above, the reddening parameter gets negative if a star emits excessively in the UV, a particular cause of such an excess being disk accretion. Therefore, the trend is suggestive of accretion disks (enhanced stellar UV) that emit thermal NIR stronger at higher accretion rates. This inference is consistent with the presence of a similar trend in the lower panel of Figure 5 (at $dM_{c0} < 0$, the NIR excess gets higher with decreasing $dM_{c0}$). Indeed, since excessive UV emission drives $dM_{c0}$ down, larger near-infrared excess at large negative values of luminosity anomaly is exactly what one would expect for accretion-disk stars in which the dust NIR emission correlates with the enhanced stellar UV emission.

In the range $dM_{c0} > 0$, the larger luminosity anomaly, the farther away from the ZAMS the younger the putative PMS stars are. Since younger systems are, at a given temperature, more luminous and typically possess more circumstellar dust, they are expected to have a stronger NIR emission. Therefore, as long as UV excess is small enough or absent, the thermal infrared emission should be stronger at higher $dM_{c0}$. The behavior of $\Delta(V-K)_{b-y}$ in the lower panel in Figure 5 at $dM_{c0} > 0$ is thus consistent with very dusty PMS stars that strongly absorb UV flux or have low disk accretion rates.

**Association with star formation regions.** Sky distribution of stars with and without near-infrared excess (Figure 6) reveals a remarkable difference between these two groups. The first group (left panel) exhibits strong association with at least two most famous regions of ongoing star formation, ρ Oph and Tau – Aur, that are centered at a distance of ~150 pc. At the same time there is nothing special about these regions in the distribution of stars in the second group (right panel). Thus, the PMS candidates identified in this paper on the basis of their NIR excess map nicely the known regions of star formation. The mapping is quite good not only in the sky but also along the line of sight: The distance distribution of the PMS candidates within the distance range 100 to 200 pc is peaked at ~150 pc, meaning that most of those stars are indeed spatially associated with the star-forming regions.

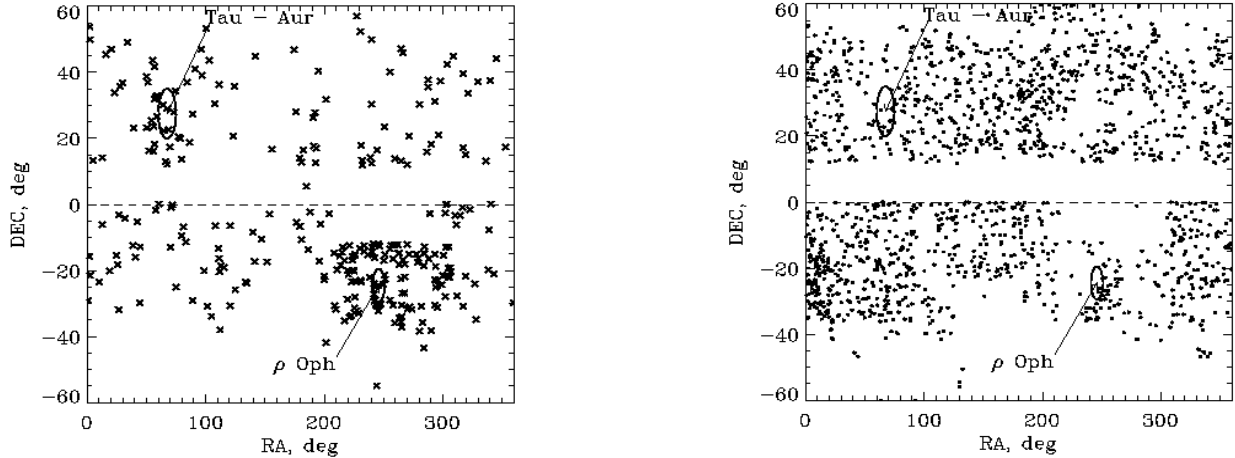

Figure 6. Sky distribution of stars within 100 to 200 pc with and without infrared excess, $\Delta(V-K)_{b-y} > 0.1$ (left) and $\Delta(V-K)_{b-y} < 0.1$ (right), respectively.

A more detailed analysis unveils many interesting features in that association. In particular, it appears that candidate PMS F stars envelope rather than get immersed into molecular clouds in the center of a star-forming region. This may be, of course, a result of an observational bias, because stars within or behind molecular clouds are hard to detect due to strong extinction. On the other hand, concentration of PMS F stars at the edges of star-forming regions may well be a real phenomenon, reflecting changes in the ISM accompanying star formation events. Further studies of the association of the candidate PMS F stars with molecular clouds in star-forming regions would certainly help to better understanding the history of these regions.